\documentclass{aa}  

\usepackage{graphicx}
\usepackage{txfonts}
\usepackage{hyperref}
\hypersetup{colorlinks=true,linkcolor=blue,citecolor=blue,filecolor=blue,urlcolor=blue}
\newcommand{\ha}{H$\alpha$~}

\newcommand{\hei}{\ion{He}{i}~}
\newcommand{\lam}{$\lambda$}
\begin{document} 

    \title{Unveiling optical signatures of outflows in accreting white dwarfs}
   \titlerunning{Optical outflows in AWDs} 
   \authorrunning{V.~A.~C\'uneo et al.}

   \author{V.~A.~C\'uneo \inst{1,2}, 
          T.~Mu\~noz-Darias \inst{1,2},
          F.~Jim\'enez-Ibarra \inst{3},
          G.~Panizo-Espinar \inst{1,2},
          J.~S\'anchez-Sierras \inst{1,2}, 
          M.~Armas~Padilla \inst{1,2},
          J.~Casares \inst{1,2},
          D.~Mata~S\'anchez \inst{1,2},
          M.~A.~P.~Torres \inst{1,2},
          F.~Vincentelli \inst{1,2},
          \and
          A.~Ambrifi \inst{1,2}
          }

   \institute{Instituto de Astrofísica de Canarias, Vía Láctea s/n, 38205 La Laguna, Tenerife, Spain\\
              \email{virginiacuneo@gmail.com}
         \and
             Departamento de Astrofísica, Universidad de La Laguna, Av. Astrofísico Francisco Sánchez s/n, 38206 La Laguna, Tenerife, Spain
         \and
             School of Physics and Astronomy, Monash University, Clayton Campus, VIC 3800, Australia \\
             }

   \date{Received 23 June 2023 / Accepted 21 August 2023}
   %\date{Received; Accepted}

% \abstract{}{}{}{}{} 
% 5 {} token are mandatory
 
  \abstract{
   Accreting white dwarfs are known to show signatures of wind-type outflows in the ultraviolet. However, at optical wavelengths, wind detections have only been reported for a few sources. We present GTC-10.4m optical spectroscopy of four accreting white dwarfs (\mbox{BZ Cam}, \mbox{V751 Cyg}, MV Lyr, and V425 Cas) observed during luminous epochs, when their optical emission is expected to be dominated by the accretion disc. Our analysis focuses on four emission lines: \ha and \hei \lam5876, \lam6678, and \lam7065. Line profiles are complex and variable on short (minutes) and long (days to weeks) timescales, with transient absorption and emission components. Among them, we detect strong blueshifted absorptions at $\gtrsim 1000$ km s$^{-1}$. These high-velocity components, present only in the blue wing of the emission lines, are observed in all four sources and could be associated with accretion disc winds. For \mbox{MV Lyr} and \mbox{V425 Cas}, these would represent the first detection of optical outflows in these objects, while in the cases of \mbox{BZ Cam} and \mbox{V751 Cyg}, the presence of outflows has been previously reported. This study suggests that, in addition to ultraviolet winds, optical outflows might also be common in accreting white dwarfs. We discuss the observational properties of these winds and their possible similarity to those detected in accreting black holes and neutrons stars.}


   \maketitle
%
%________________________________________________________________

\section{Introduction}
Accreting white dwarfs (AWDs; also known as cataclysmic variables) are binary systems typically composed of a non-degenerated companion star that is filling its Roche lobe and transferring matter onto a white dwarf through the inner Lagrangian point. In non-magnetic systems, the transferred material forms an accretion disc around the white dwarf, the relatively low magnetic field of which is not strong enough to disrupt it. Among the population of AWDs, nova-like systems are quasi-persistently accreting at high rates \mbox{($\dot{M} \sim 10^{-8}$ M$_{\odot}$ yr$^{-1}$)} and have bright accretion discs. Nova-like AWDs can also show long-term variations in their optical light curves and exhibit sporadic low-activity periods characterised by drops of $\gtrsim$ 1 mag \citep{King1998}. 

Wind-type outflow features have been observed in AWDs. These are generally detected in ultraviolet resonant lines, such as \mbox{\ion{C}{iv} \lam\lam1548-1552}, \mbox{\ion{Si}{iv} \lam\lam1393-1402}, and \ion{N}{iv} \lam\lam1238-1242 \citep[e.g.][]{Drew1997,Prinja2000a}. The strong ultraviolet emission produced by AWDs is generally believed to power these winds via the line-driven mechanism \citep[e.g.][]{Froning2005}. However, this is still a matter of debate (see \citealt{Drew2000} and references therein). These outflows have shown velocities of up to $\sim$5000 km s$^{-1}$. Given that they have only been observed in non-eclipsing systems ---that is, AWDs seen at low orbital inclinations ($\lesssim 60^{\circ}$)---, a bipolar wind geometry is generally favoured \citep[e.g.][]{Honeycutt1986,Drew1987}.  

A handful of the AWDs with strong ultraviolet winds have also shown optical wind features, such as P-Cygni profiles (a blueshifted absorption component superimposed on a broad emission feature; \citealt{Castor1979}) and broad emission line wings, both overlapping the emission lines formed in the accretion disc. These usually include the Balmer lines and \hei triplet transitions (e.g. \hei \lam5876), but not the less intense \hei singlet transitions \citep[such as \hei \lam6678; e.g.][]{Ringwald1998,Patterson2001,Kafka2004,Honeycutt2013}. The presence of optical winds has also been suggested based on the detection of  more indirect signatures. For example, \citet{Murray1996} showed that single-peaked emission lines might result from an accelerating accretion disc wind in high-luminosity AWDs. Likewise, \citet{Knigge1998} proposed that the Balmer jump in emission detected in the system UX~UMa is evidence of the presence of an outflow.

Furthermore, it has been proposed that the visibility of wind features in AWDs has an orbital dependence (see \citealt{Prinja2004} for an example of ultraviolet winds, and \citealt{Kafka2009} for optical winds). It is worth noting that qualitatively similar optical wind features have been found in accreting black holes \citep[e.g.][]{Munoz-Darias2016,Munoz-Darias2019,Cuneo2020a} and neutron stars \citep[e.g.][]{Munoz-Darias2020} using high-signal-to-noise-ratio (S/N) spectroscopy. These are detected in several transitions (including the Balmer series and \hei triplet transitions, but also the \hei singlet transitions) and point towards wind velocities of up to a few thousand \mbox{km s$^{-1}$}, comparable to those measured in AWDs.  

Optical spectroscopic studies of samples of nova-like AWDs are relatively limited and generally taken with low-to-moderate spectral resolution and S/N \citep[e.g.][]{Kafka2004,Aungwerojwit2005,Rodriguez-Gil2007b}. In order to systematically search for optical outflows in AWDs, we started a multi-epoch observational campaign using the 10.4m Gran Telescopio Canarias (GTC). In this paper, we present the first results obtained for four well-known systems that show complex and variable absorption components: BZ Cam, \mbox{V751 Cyg}, MV Lyr, and \mbox{V425 Cas}.

%-------------------------------------------------------------------
\section{The sources}
\label{sources}
\begin{figure}
 \includegraphics[trim=3mm 0mm 7mm 0mm,width=0.95\columnwidth]{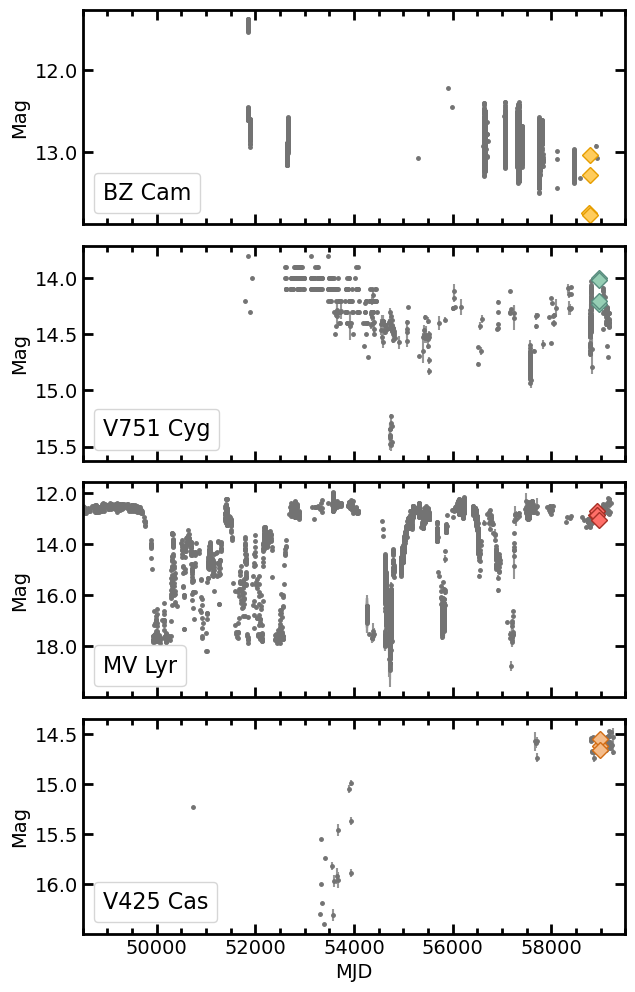}
 \caption{Light curves from AAVSO data combining V and Clear V-band magnitudes. The diamonds denote the r magnitude in our observing epochs.}
 \label{lc}
\end{figure}

Table
\ref{sample} contains the main properties of the sample relevant for our study.  Figure \ref{lc} shows the light curves from American Association of Variable Star Observers (AAVSO) data in the V and Clear V bands, where we mark our observing epochs (see also Table \ref{log}). A description of the four systems of our sample is provided in the following paragraphs.

%------------------------------------------
\begin{table}
\caption{General properties of the AWDs sample.}             
\label{sample}      
\centering        
\resizebox{\columnwidth}{!}{
\begin{tabular}{l c c c}
\hline\hline            
\rule{0pt}{2.5ex}Source & Orbital period (h) & Inclination ($^{\circ}$) & References \\
\\[-2ex]
\hline     
   \rule{0pt}{2.3ex}BZ Cam & 3.685$\pm$0.001 & 12--40 & 1, 2 \\   
   \rule{0pt}{2ex}V751 Cyg & 3.46714$\pm$0.00002 & $<$ 50 & 3, 4 \\ 
   \rule{0pt}{2ex}MV Lyr & 3.19$\pm$0.01 & 10--13; 7$\pm$1 & 5, 6 \\ 
   \rule{0pt}{2ex}V425 Cas & 3.59$\pm$0.01 & 25$\pm$9 & 7, 8 \\ 
   \hline   
\end{tabular}} 

\tablebib{(1) \citet{Ringwald1998}; (2) \citet{Honeycutt2013}; (3) \citet{Greiner1999}; (4) \citet{Patterson2001}; (5) \citet{Skillman1995}; (6) \citet{Linnell2005}; (7) \citet{Shafter1982}; (8) \citet{Ritter2003}.}
\end{table}
%------------------------------------------

\begin{table*}
\caption{Observing log.}            
\label{log}      
\centering   
\begin{tabular}{@{\extracolsep{\fill}} l@{\hskip 0.05cm} c@{\hskip 0.05cm} c@{\hskip 0.25cm} c@{\hskip 0.2cm} c@{\hskip 0.05cm} c@{\hskip 0.5cm} l@{\hskip 0.05cm} c@{\hskip 0.05cm} c@{\hskip 0.25cm} c@{\hskip 0.2cm} c@{\hskip 0.05cm}}     
\hline\hline       
\rule{0pt}{2.5ex}Source & Epoch & Date (hh:mm:ss) & MJD & r (mag) & & Source & Epoch & Date (hh:mm:ss) & MJD & r (mag) \\
\\[-2ex]
\hline
\rule{0pt}{2.5ex}\textbf{BZ Cam} & 1a & 09/10/19 (05:16:23) & 58765.2197  & 13.75 &  & \textbf{MV Lyr} & 1a & 08/03/20 (06:42:57) & 58916.2798 & 12.86  \\
\rule{0pt}{2.5ex}       & 1b & 09/10/19 (05:18:47) & 58765.2214 &  &  &  & 1b & 08/03/20 (06:44:10) & 58916.2807 &  \\
\rule{0pt}{2.5ex}       & 1c & 09/10/19 (05:21:10) & 58765.2230 &  &  &  & 1c & 08/03/20 (06:45:24) & 58916.2815 &  \\
\rule{0pt}{2.5ex}       & 1d & 09/10/19 (05:23:33) & 58765.2247 &  &  &  & 1d & 08/03/20 (06:46:38) & 58916.2824 &  \\
\cline{2-5} \cline{8-11}
\rule{0pt}{2.5ex}       & 2a & 20/10/19 (03:57:07) & 58776.1647 & 13.76  &  &  & 2a & 09/03/20 (06:26:31) & 58917.2684 & 12.69 \\
\rule{0pt}{2.5ex}       & 2b & 20/10/19 (04:00:36) & 58776.1671 &  &  &  & 2b & 09/03/20 (06:28:54) & 58917.2701 &  \\
\rule{0pt}{2.5ex}       & 2c & 20/10/19 (04:02:40) & 58776.1685 &  &  &  & 2c & 09/03/20 (06:31:21) & 58917.2718 &  \\
\rule{0pt}{2.5ex}       & 2d & 20/10/19 (04:04:03) & 58776.1695 &  &  &  & 2d & 09/03/20 (06:33:47) & 58917.2735 &  \\
\rule{0pt}{2.5ex}       & 2e & 20/10/19 (04:05:28) & 58776.1705 &  &  &  &  &  &  &  \\
\cline{2-5} \cline{8-11}
\rule{0pt}{2.5ex}       & 3a & 22/10/19 (01:53:44) & 58778.0790 & 13.04 &  &  & 3a & 10/03/20 (06:19:14) & 58918.2634 & 12.86 \\
\rule{0pt}{2.5ex}       & 3b & 22/10/19 (01:56:07) & 58778.0806 &  &  &  & 3b & 10/03/20 (06:21:40) & 58918.2650 &  \\
\rule{0pt}{2.5ex}       & 3c & 22/10/19 (01:58:31) & 58778.0823 &  &  &  & 3c & 10/03/20 (06:24:05) & 58918.2667 &  \\
\rule{0pt}{2.5ex}       & 3d & 22/10/19 (02:00:54) & 58778.0840 &  &  &  & 3d & 10/03/20 (06:26:28) & 58918.2684 &  \\
\cline{8-11}
\rule{0pt}{2.5ex}       & 3e & 22/10/19 (02:47:12) & 58778.1161 & 13.27  &  &  & 4a & 03/05/20 (03:44:53) & 58972.1562 & 13.04  \\
\rule{0pt}{2.5ex}       & 3f & 22/10/19 (02:49:36) & 58778.1178 &  &  &  & 4b & 03/05/20 (03:46:16) & 58972.1571 &  \\
\rule{0pt}{2.5ex}       & 3g & 22/10/19 (02:51:59) & 58778.1194 &  &  &  & 4c & 03/05/20 (03:47:40) & 58972.1581 &  \\
\rule{0pt}{2.5ex}       & 3h & 22/10/19 (02:54:23) & 58778.1211 &  &  &  & 4d & 03/05/20 (03:49:03) & 58972.1591 &  \\
\hline
\rule{0pt}{2.5ex}\textbf{V751 Cyg} & 1a & 03/05/20 (05:20:43) & 58972.2227 & 14.00 &  & \textbf{V425 Cas} & 1a & 18/05/20 (04:47:22) & 58987.1996 & 14.63  \\
\rule{0pt}{2.5ex}       & 1b & 03/05/20 (05:23:36) & 58972.2247 &  &  &  & 1b & 18/05/20 (04:51:06) & 58987.2021 &  \\
\rule{0pt}{2.5ex}       & 1c & 03/05/20 (05:26:30) & 58972.2267 &  &  &  & 1c & 18/05/20 (04:54:49) & 58987.2047 &  \\
\rule{0pt}{2.5ex}       & 1d & 03/05/20 (05:29:23) & 58972.2287 &  &  &  & 1d & 18/05/20 (04:58:33) & 58987.2073 &  \\
\cline{2-5} \cline{8-11}
\rule{0pt}{2.5ex}       & 2a & 04/05/20 (05:16:08) & 58973.2195 & 14.23  &  &  & 2a & 20/05/20 (05:07:15) & 58989.2134 & 14.55  \\
\rule{0pt}{2.5ex}       & 2b & 04/05/20 (05:19:02) & 58973.2216 &  &  &  & 2b & 20/05/20 (05:10:58) & 58989.2160 &  \\
\rule{0pt}{2.5ex}       & 2c & 04/05/20 (05:21:55) & 58973.2236 &  &  &  & 2c & 20/05/20 (05:14:42) & 58989.2185 &  \\
\rule{0pt}{2.5ex}       & 2d & 04/05/20 (05:24:48) & 58973.2256 &  &  &  & 2d & 20/05/20 (05:18:25) & 58989.2211 &  \\
\cline{2-5} \cline{8-11}
\rule{0pt}{2.5ex}       & 3a & 05/05/20 (05:18:36) & 58974.2213 & 14.21  &  &  & 3a & 22/05/20 (04:46:53) & 58991.1992 & 14.67  \\
\rule{0pt}{2.5ex}       & 3b & 05/05/20 (05:21:30) & 58974.2233 &  &  &  & 3b & 22/05/20 (04:50:36) & 58991.2018 &  \\
\rule{0pt}{2.5ex}       & 3c & 05/05/20 (05:24:23) & 58974.2253 &  &  &  & 3c & 22/05/20 (04:54:20) & 58991.2044 &  \\
\rule{0pt}{2.5ex}       & 3d & 05/05/20 (05:27:17) & 58974.2273 &  &  &  & 3d & 22/05/20 (04:58:03) & 58991.2070 &  \\
\cline{2-5}
\rule{0pt}{2.5ex}       & 4a & 06/05/20 (05:11:35) & 58975.2164 & 14.01 &  &  &  &  &  &  \\
\rule{0pt}{2.5ex}       & 4b & 06/05/20 (05:14:28) & 58975.2184 &  &  &  &  &  &  &  \\
\rule{0pt}{2.5ex}       & 4c & 06/05/20 (05:17:21) & 58975.2204 &  &  &  &  &  &  &  \\
\rule{0pt}{2.5ex}       & 4d & 06/05/20 (05:20:15) & 58975.2224 &  &  &  &  &  &  &  \\
\hline                    
\end{tabular}
\end{table*}

\paragraph{BZ Cam} is one of the most studied AWDs. The system is surrounded by an optically resolved bow-shock nebula, photo-ionised by the AWD \citep[e.g.][]{Greiner2001}. The V-band brightness of the system usually varies between 13.5 mag and \mbox{11.4 mag} (see \mbox{Fig. \ref{lc}).} BZ Cam is known to have strong and variable outflows (evident through P-Cygni profiles), which were first detected in ultraviolet low-resolution spectra provided by the \textit{International Ultraviolet Explorer} \citep[\textit{IUE;}][]{Woods1990,Griffith1995}. \citet[][]{Prinja2000a} studied the outflow's properties using \textit{Hubble Space Telescope} (HST) high-resolution spectra and measured the blue edges of the blueshifted absorption features in the \mbox{P-Cygni} profiles ---which is usually taken as a good proxy for the terminal velocity of the wind--- of C, Si, and N lines. This resulted in wind velocities of up to \mbox{$\sim$5000 km s$^{-1}$}. Furthermore, BZ Cam is one of the few AWDs with outflows detected at optical wavelengths \citep[][]{Patterson1996,Ringwald1998,Honeycutt2013}. These optical outflows, revealed by strong \mbox{P-Cygni} profiles in H and He emission lines, reach velocities of up to \mbox{$\sim$3000 km s$^{-1}$}. In addition to the \mbox{P-Cygni} profiles, both \citet[][]{Patterson1996} and \citet[][]{Ringwald1998} observed broad red emission wings in H$\alpha$ up to \mbox{2400 km s$^{-1}$}, features typically interpreted as evidence of outflows. \citet[][]{Ringwald1998} estimated an orbital inclination of between 12$^{\circ}$ and 40$^{\circ}$, which is consistent with the lack of eclipses. 

\paragraph{V751 Cyg} has a typical brightness of $\sim$14-15 mag in the \mbox{V band} (see Fig. \ref{lc}) but has been reported to show lower luminosity states down to V$\sim$17.8 mag \citep[see e.g.][]{Greiner1999b}. A shallow P-Cygni profile (\ion{C}{iv} \lam1550) is the only reported detection of ultraviolet outflows \citep{Zellem2009}. Additionally, \citet{Patterson2001} detected transient P-Cygni profiles in the Balmer and \hei \lam5876 emission lines with terminal velocities of up to 2500 km s$^{-1}$. Based on the equivalent width (EW) of several ultraviolet spectral lines, \citet[][]{Greiner1999} estimated an orbital inclination of $<$ 50$^{\circ}$.

\paragraph{MV Lyr} spends most of the time in a bright state, at \mbox{V$\sim$12.5 mag} (see Fig. \ref{lc}), but it also shows sporadic low states, when its brightness can drop down to \mbox{$\sim$19 mag.} Although compelling outflow signatures have not yet been detected in the optical, \citet[][]{Skillman1995} observed a red tail in the emission profile of H$\alpha$, with velocities of up to \mbox{1300 km s$^{-1}$}. Likewise, \citet[][]{Linnell2005} proposed the presence of a low-velocity (\mbox{$\sim$165 km s$^{-1}$}) ultraviolet wind through the detection of blueshifted (C, Si and He) absorption lines in HST spectra. \citet[][]{Skillman1995} estimated a very low orbital inclination in the range of \mbox{10$^{\circ}$--13$^{\circ}$}, while \citet[][]{Linnell2005} proposed 7$\pm$1$^{\circ}$. 

\paragraph{V425 Cas} has been recurrently observed at V$\sim$14.5 (Fig. \ref{lc}), although it has been seen to drop down to V$\sim$18 during low states \citep{Wenzel1987}. To date, no wind-related features have been published at either optical or ultraviolet wavelengths \citep[e.g.][]{Szkody1985,Mizusawa2010}. An orbital inclination of 25$\pm$9$^{\circ}$ was reported by \citet[][with reference to \citealp{Shafter1983}]{Ritter2003}.

%-------------------------------------------------------------------
\section{Observations}
\label{obs}
We observed the sources during 2019 and 2020 using the Optical System for Imaging and low-Intermediate-Resolution Integrated Spectroscopy \citep[OSIRIS;][]{Cepa2000}, which is attached to GTC at the Observatorio del Roque de los Muchachos in La Palma, Spain. We observed each source on three or four different nights over a timespan of 4 to 57 days, depending on the target. For each epoch, we obtained at least four spectra with a cadence of $\sim$2--4 minutes. The observing log is detailed in \mbox{Table \ref{log}}. The exposure times ranged between 50 and 200 s. We used the R2500R (5575--7685 \AA) grism and a slit-width of 1.2 arcsec. This configuration resulted in slit-limited velocity resolutions of 200--240 km s$^{-1}$, as derived from the full width at half-maximum (FWHM) of the \ion{O}{i} $\lambda$6300 emission sky-line. The reduction, extraction, and wavelength calibration of the spectra were carried out with \textsc{iraf}\footnote{IRAF is distributed by the National Optical Astronomy Observatory, which is operated by the Association of Universities for Research in Astronomy, Inc. under contract to the National Science Foundation.}. We achieved S/N values ranging from 100 to 350.

\begin{figure*}
 \includegraphics[trim=0mm 3mm 0mm 0mm,clip,width=\textwidth]{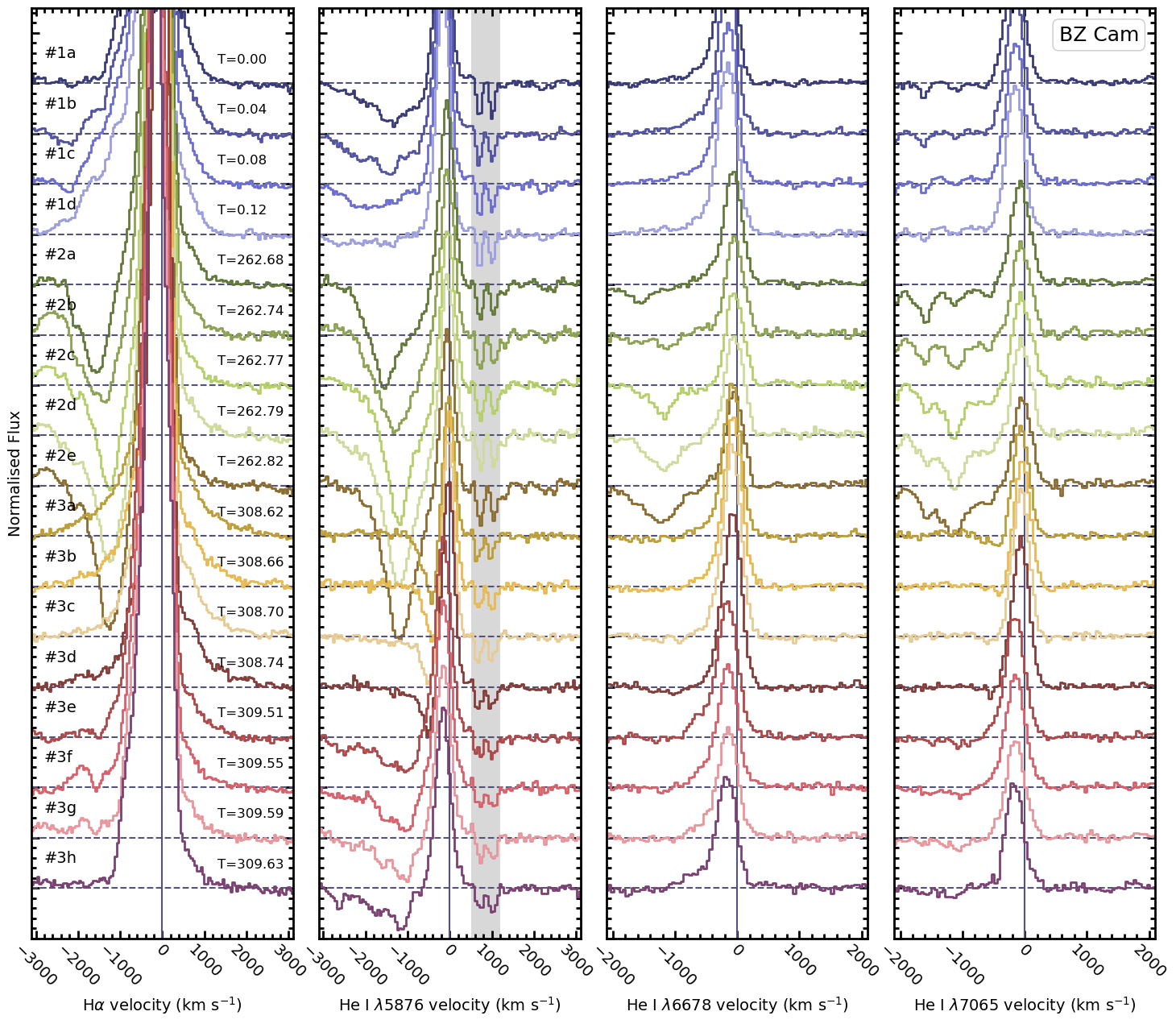}
 \caption{Evolution of the H$\alpha$, \ion{He}{i} $\lambda$5876, $\lambda$6678, and $\lambda$7065 emission line profiles for BZ Cam along three different observing epochs (from top to bottom, respectively). All spectra are normalised to their continuum level and separated by an offset of 0.1 in the vertical axis for clarity. The time since our first observation, T, is given in hours. The grey-shaded band indicates contamination by interstellar features.}
 \label{bz_cam}
\end{figure*}

We derived optical magnitudes from the acquisition images of each observing run. Flux calibration was performed against field stars catalogued in PanSTARRS \citep{Magnier2020} and using routines based on \textsc{astropy-photutils}  \citep{Bradley2019}. We report an averaged magnitude for each night in Table \ref{log} (for epoch \#3 of \mbox{BZ Cam} we report one average magnitude per observing run). The uncertainties were two percent or better.

%--------------------------------------------------------------------
\section{Analysis and results}
\label{ana}
\begin{figure*}
 \includegraphics[trim=0mm 3mm 0mm 0mm,clip,width=\textwidth]{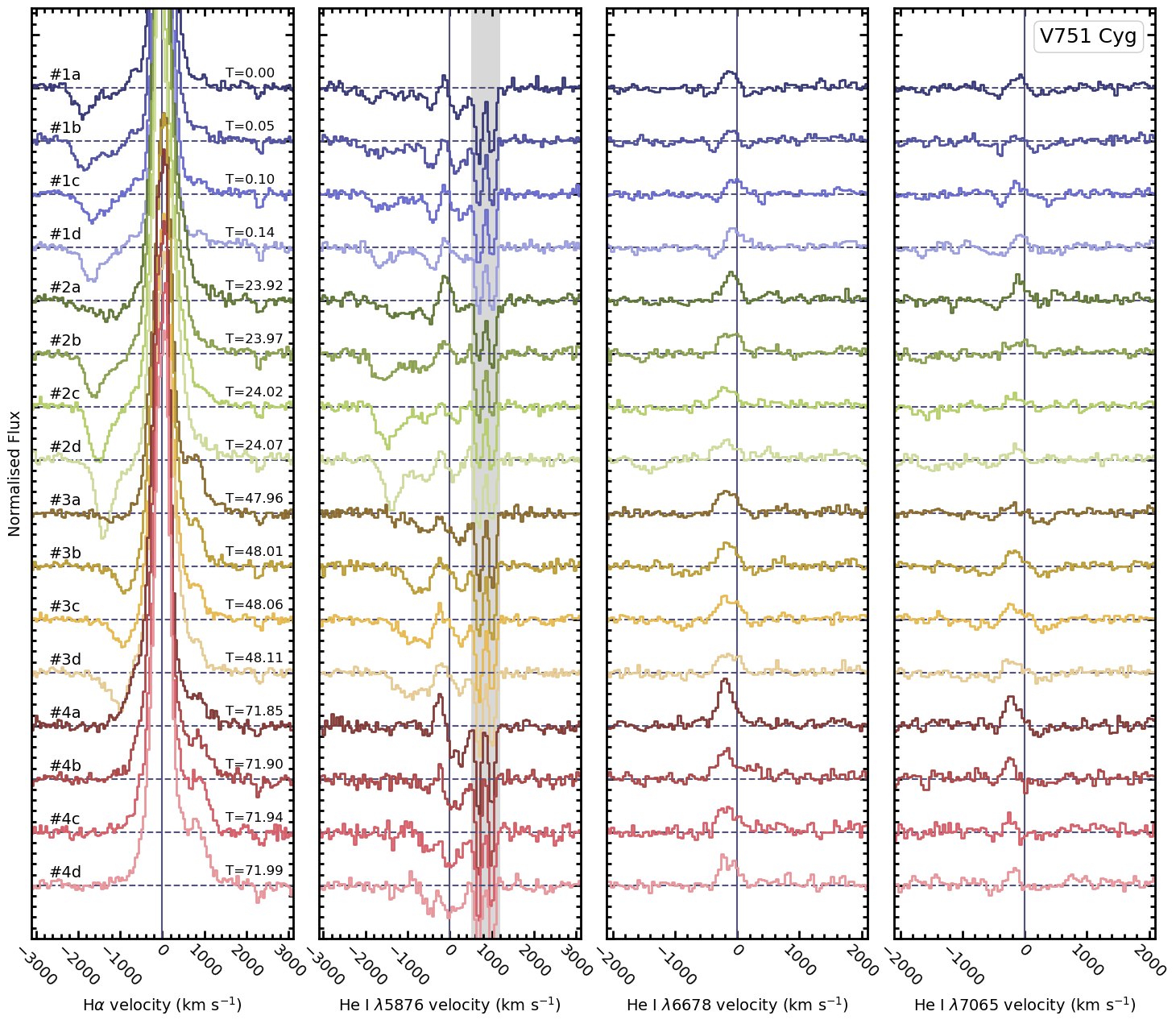}
 \caption{Evolution of the H$\alpha$, \ion{He}{i} $\lambda$5876, $\lambda$6678, and $\lambda$7065 emission line profiles for V751 Cyg along four different observing epochs (from top to bottom, respectively). All spectra are normalised to their continuum level and separated by an offset of 0.1 in the vertical axis for clarity. The time since our first observation, T, is given in hours. The grey-shaded band indicates contamination by interstellar features.}
 \label{v751_cyg}
\end{figure*}

\begin{figure*}
 \includegraphics[trim=0mm 3mm 0mm 0mm,clip,width=\textwidth]{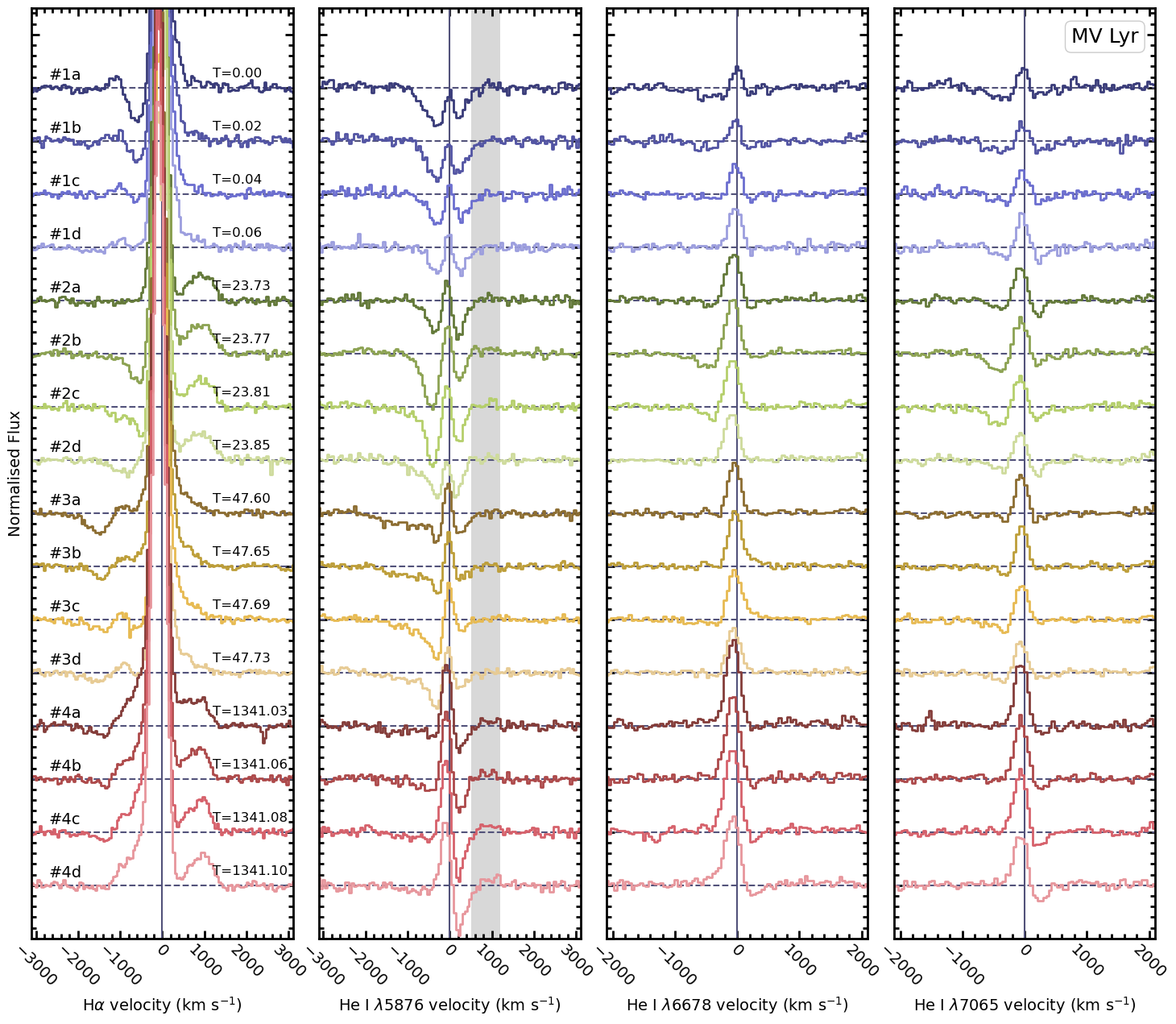}
 \caption{Evolution of the H$\alpha$, \ion{He}{i} $\lambda$5876, $\lambda$6678 and $\lambda$7065 emission line profiles for MV Lyr along four different observing epochs (from top to bottom, respectively). All spectra are normalised to their continuum level and separated by an offset of 0.1 in the vertical axis for clarity. The time since our first observation, T, is given in hours. The grey shaded band indicates contamination by interstellar features.}
 \label{mv_lyr}
\end{figure*}

\begin{figure*}
 \includegraphics[trim=0mm 3mm 0mm 0mm,clip,width=\textwidth]{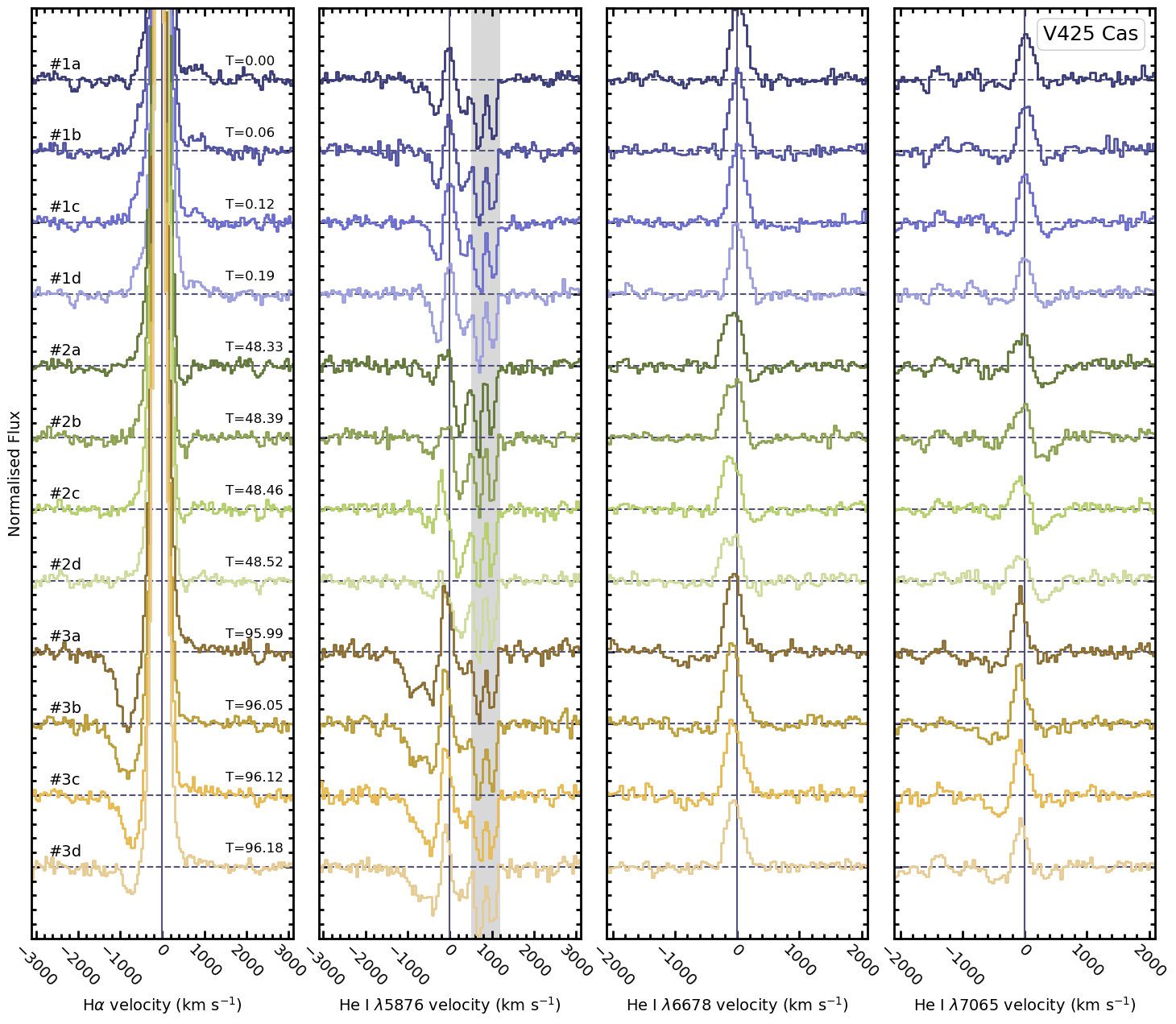}
 \caption{Evolution of the H$\alpha$, \ion{He}{i} $\lambda$5876, $\lambda$6678, and $\lambda$7065 emission line profiles for V425 Cas along three different observing epochs (from top to bottom, respectively). All spectra are normalised to their continuum level and separated by an offset of 0.1 in the vertical axis for clarity. The time since our first observation, T, is given in hours. The grey-shaded band indicates contamination by interstellar features.}
 \label{v425_cas}
\end{figure*}

We analysed the optical spectra using \textsc{molly} and custom software under \textsc{python} 3.7. \ha and the main \hei transitions are present in emission in all four sources, as expected for systems with bright accretion discs. We focused our analysis on transitions that are notorious wind tracers: H$\alpha$ and the triplets \hei \lam5876 and \lam7065, but also on the singlet transition \hei \lam6678. We carefully normalised each line by fitting the adjacent continuum with a first-order polynomial. Both \ha and \hei lines exhibit significant variability across the observing campaign. Their evolution can be seen in Figs. \ref{bz_cam}, \ref{v751_cyg}, \ref{mv_lyr}, and \ref{v425_cas}.

%---------------------------------------
\subsection{Evolution of the main emission lines}
\label{ana_evol}
\begin{figure*}
 \centering
 \includegraphics[trim=3mm 3mm 0mm 0mm,width=0.88\textwidth]{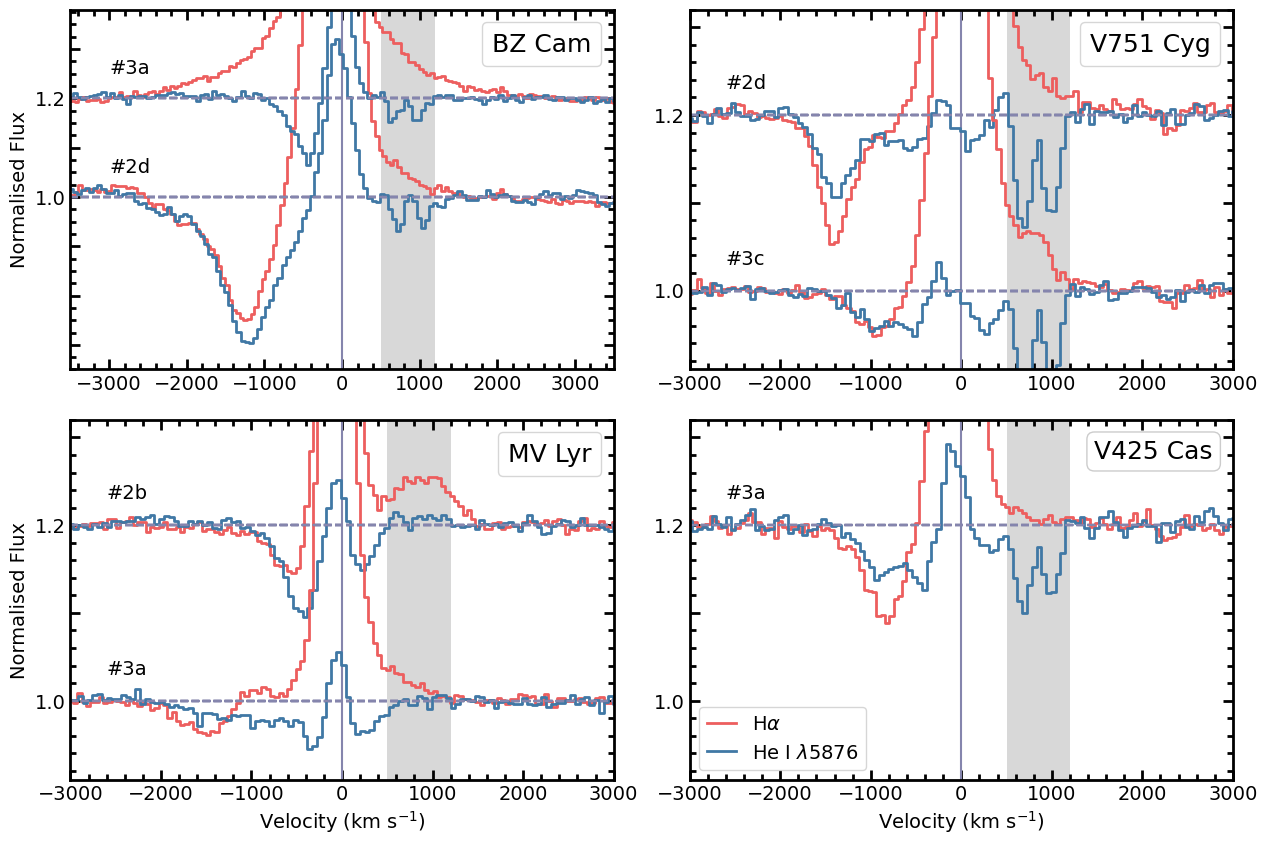}
 \caption{Comparison between \ha (red) and \hei \lam5876 (blue) line profiles for different epochs of the four sources. An offset of 0.2 in the \textit{y}-axis was applied for clarity. The grey-shaded band indicates contamination of the \hei \lam5876 line profile by interstellar features.}
 \label{overlap}
\end{figure*}

Optical, double-peaked emission lines are commonly found in compact binaries and interpreted as arising in the accretion disc \citep{Smak1969}. These can also be seen as single-peaked emission components in
systems with low orbital inclination (as is presumably the case for the four sources studied here). On the other hand, (intrinsic) single-peaked emission lines are also often detected during (presumably) disc-dominated phases (see e.g. \citealt{Matthews2015} and references therein). However, as shown in \mbox{Figs. \ref{bz_cam}}, \ref{v751_cyg}, \ref{mv_lyr}, and \ref{v425_cas}, the optical lines of AWDs can be significantly more complex than single- or double-peaked emission components. In addition to the possible presence of \mbox{wind-related} features (e.g. \mbox{P-Cygni} profiles), absorptions that shift in velocity throughout the orbit ---producing variable components at both the red and blue side of the emission peak--- are sometimes observed in AWDs \citep[e.g.][]{Casares1996,Hoard2000,Kara2023}. Our study lacks the orbital coverage to properly single out these components. However, hints indicating the possible presence of underlying absorption features in the range of $\sim -1000$ to \mbox{1000 km s$^{-1}$} can be seen in some of the sources, particularly in the \hei transitions. Figure \ref{overlap} (see also \mbox{Section \ref{velocities}}) shows an overlap of the \ha and \hei \lam5876 emission lines  for the four sources in our sample. As an example of the multiple components that we detect, the \hei \lam5876 transition in epoch \#3c of V751 Cyg (top right panel) presents a main emission component that seems to be embedded in a broad absorption, which is observed as both redshifted and blueshifted absorption features that reach velocities below 1000 km s$^{-1}$. An additional blueshifted absorption component ---that overlaps its analogue in H$\alpha$--- reaches velocities higher than 1000 km s$^{-1}$. Hereafter, we refer to the absorption features in the range of $\sim -1000$ to 1000 km s$^{-1}$ as low-velocity absorptions, and those with velocities \mbox{$\gtrsim 1000$ km s$^{-1}$} as high-velocity absorptions. The same phenomenology is observed in other epochs and sources (e.g. epochs \#3a of \mbox{MV Lyr} and V425 Cas). Likewise, steady, broad absorption features overlapping the main emission profiles have also been observed in the Balmer transitions during active states of accreting black holes and neutron stars [i.e. X-ray binaries (XRBs); e.g. \citealt[][]{Casares1995,Callanan1995,Soria2000a,Jimenez-Ibarra2019a}]. For all these reasons, any absorption feature found in the data should be interpreted with caution.

\paragraph{BZ Cam:} The spectral evolution of BZ~Cam is depicted in \mbox{Fig. \ref{bz_cam}}. The profile of all the lines is single-peaked. We noticed that the core of \ha is saturated in epoch \#1; however, this is not a major issue given that we focus our analysis on the wings of the line. Strong high-velocity blueshifted absorptions overlapping the central emission feature are evident in our spectra (more conspicuous in epoch \#2) in all the transitions analysed. A variable low-velocity broad absorption component overlapping both the central emission and the blueshifted high-velocity absorption component is often also seen in the spectra (e.g. \hei \lam5876 in epoch \#3). In addition to the absorption features, both the blue and red wings of the emission lines (e.g. \ha in epochs \#1 and \#3) show clear asymmetries. 

\paragraph{V751 Cyg:} We display the \ha and \hei emission lines of \mbox{V751 Cyg} in Fig. \ref{v751_cyg}. All the emission features have a single-peaked profile. The line profiles ---in particular \ha and \hei \lam5876--- exhibit an overlap between the main emission core and several absorption components. In some cases, the presence of a variable low-velocity broad absorption component (producing red and blue absorptions) and a high-velocity blueshifted component can be resolved (e.g. \hei \lam5876 in most epochs; see also Fig. \ref{overlap}). The red wing of \ha shows additional asymmetries, particularly in epochs \#3 and \#4. 

\paragraph{MV Lyr:} This source also shows single-peaked line profiles (Fig. \ref{mv_lyr}), with the main emission component blended with several absorption components. Low-velocity variable absorption features are observed in all the lines. An additional high-velocity blueshifted absorption component is particularly conspicuous in epoch \#3 (\ha and \hei \lam5876; see also Fig. \ref{overlap}). Additional asymmetries are evident in the red wing of H$\alpha$, particularly in epochs \#2 and \#4.

\paragraph{V425 Cas:} Figure \ref{v425_cas} shows the evolution of the \ha and \hei emission lines for V425 Cas. Most of the lines exhibit a single-peak profile. However, a double peak can be noticed in \hei \lam6678 in epoch \#2d. The different absorption components detected for the previous sources are also visible in the spectra of V425 Cas. While the low-velocity absorption component is observed in the three \hei lines, the high-velocity blueshifted component is evident in \ha and \hei \lam5876 in epoch \#3 (see also Fig. \ref{overlap}).

%---------------------------------------
\subsection{\texorpdfstring{H$\alpha$}{Halpha} excesses diagnostic diagram}
\label{wings_diagram}

\begin{figure*}
 \centering
 \includegraphics[trim=22mm 3mm 0mm 0mm,width=0.88\textwidth]{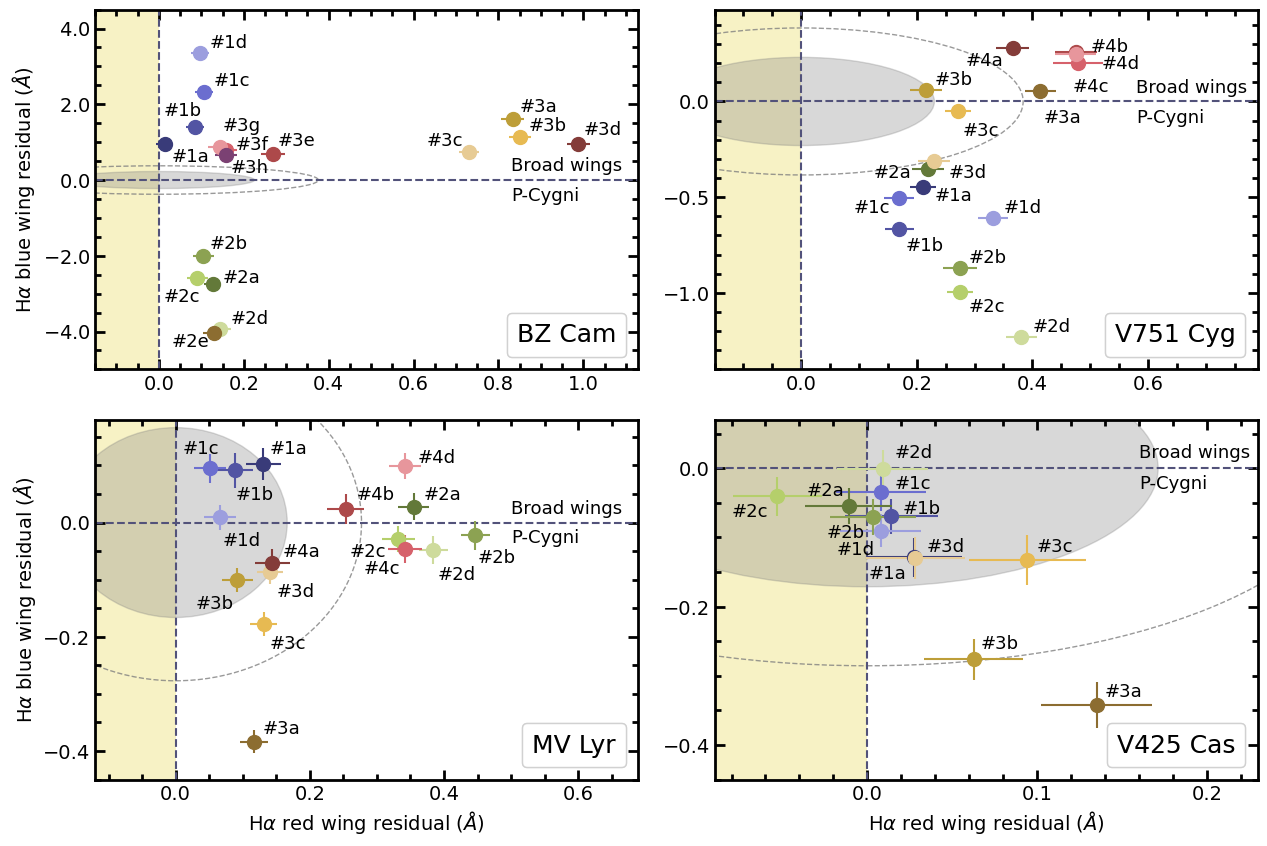}
 \caption{Excesses diagnostic diagram of the \ha wings of the four sources. In all cases, we chose a uniform spectral region \mbox{($\pm$1000-2500 km s$^{-1}$) to estimate the residuals}. The colours follow the same pattern as in Figs. \ref{bz_cam}-\ref{v425_cas}. The filled grey and dashed empty ellipses denote the 3$\sigma$ and 5$\sigma$ significance regions, respectively.}
 \label{wings}
\end{figure*}

In order to further investigate the origin of the complex emission line profiles described above, we computed the excesses diagnostic diagram (see \citealt{MataSanchez2018, Munoz-Darias2019,Panizo-Espinar2021} for details). Although the origin of the low-velocity absorption components in AWDs is unclear, the high-velocity features resemble those observed in the ultraviolet in the same systems, and the optical ones associated with winds in XRBs. Therefore, we focus our analysis on the high-velocity components. As is usually the case, we computed the excesses diagnostic diagram only for H${\alpha}$, given that the nearby Na-doublet complicates the analysis of \ion{He}{i} $\lambda$5876, and \ion{He}{i} $\lambda$6678 and $\lambda$7065 are significantly less intense. We first fitted a Gaussian profile to the central emission feature, masking the line wings. We then subtracted the fit from the spectrum and finally estimated the EW of the residuals within predefined velocity masks placed on both sides of the emission line. The significance of the residuals can be evaluated by computing EW measurements in nearby continuum regions (see \citealt{Panizo-Espinar2021}). Figure \ref{wings} shows the red wing versus the blue wing residuals for each source. Each excesses diagnostic diagram is divided in four quadrants, with two of them consistent with the presence of wind signatures: positive blue and red residuals (top-right quadrant) imply emission excesses (i.e. symmetric or asymmetric emission line wings), while blue negative residuals combined with red excesses (bottom-right) are consistent with P-Cygni profiles. To minimise the impact of the low-velocity features in the diagram, we computed the EW of the residuals using the velocity masks that better account for the higher velocity features in the four sources: \mbox{$\pm$1000-2500 km s$^{-1}$}. This approach is conservative, and is designed to limit the impact of the aforementioned broad absorption features, either static (i.e. affecting both the blue and red wings of the profile) or dynamic (e.g. shifting in velocity, from blue to red and vice versa, following the orbit of the binary). The  diagrams in Fig. \ref{wings} show that significant, high-velocity features are mostly located in the top- and bottom-right quadrants, including several $>$ 5$\sigma$ significant residuals. The only exception might be V425 Cas, for which ---using this approach--- only two observations are consistent with the presence of winds. However, two clearly distinct blueshifted absorption features are also evident in this system (see bottom right panel of \mbox{Fig. \ref{overlap}}), although at lower velocities. We note that if we use a lower velocity mask for this source (\mbox{$\pm$400-1500 km s$^{-1}$}), we obtain similar results to those obtained for the other targets (\mbox{Fig. \ref{Ha_diag_V425Cas}}). This suggests that V425 Cas shows the same phenomenology, but at lower velocities.

Overall, the systematic presence of excesses in the right quadrants and not in the left ones (i.e. no redshifted absorptions as compared with a large number of blueshifted residuals) is consistent with the idea that at least the high-velocity features (\mbox{$\gtrsim$ 1000 km s$^{-1}$}) reflect the presence of outflows in these systems. We note that the limited coverage of our data prevents us from further testing the origin of these components, including their possible dependence on orbital phase.

%---------------------------------------
\subsection{Disc-wind signatures}
\label{velocities}
As described above, the high-velocity (\mbox{$\gtrsim 1000$ km s$^{-1}$}) \mbox{blueshifted} absorption components can be tentatively associated with signatures of wind-type outflows. In these cases, we fitted a Gaussian profile to the absorption feature and estimated the wind velocity as 3$\sigma$ from the centre of the fitted Gaussian (i.e. where the blue edge meets the continuum). Hereafter, we quote this value as the wind velocity. In addition, the \ha excesses diagnostic diagram reveals significant broad emission line wings, which are also a typical feature of outflowing objects. Taking all of the above into consideration, spectral features consistent with wind-type outflows are found in the four sources: 

\paragraph{BZ~Cam:} The excesses diagnostic diagram in the top-left panel of Fig. \ref{wings} is consistent with the presence of wind in all epochs. In particular, epoch \#2 shows massive P-Cygni-like absorption features (see also Fig. \ref{bz_cam}) reaching velocities of $\sim$2400 km s$^{-1}$ (epoch \#2a). These absorption components are present even in \hei \lam6678, a transition in which wind-type features have not previously been observed for this object \citep[e.g.][]{Kafka2004,Honeycutt2013}. In the top-left panel of Fig. \ref{overlap} we show how the blue-edge velocity of the high-velocity blueshifted absorption features of \ha and \hei\lam5856 matches perfectly (epoch \#2d). In the same figure, we also show how the blue edge of \ha in epoch \#3a  matches that of the absorption in epoch \#2d. A similar behaviour has been observed in the XRBs that show the strongest optical winds (e.g. \mbox{Fig. 4} in \citealt{Munoz-Darias2018}).

\paragraph{V751~Cyg:} The excesses diagnostic diagram shown in the top-right panel of \mbox{Fig. \ref{wings}} suggests the presence of wind features in all the epochs studied. In particular, epochs \#1, \#2, and \#3 present \mbox{P-Cygni-like} absorption features (Fig. \ref{v751_cyg}) that reach velocities of \mbox{$\sim$2500 km s$^{-1}$} (epoch \#2a). In the top-right panel of \mbox{Fig. \ref{overlap}}, we show, as an example, the superposition of \ha and \hei \lam5876 for epochs \#2d and \#3c. We observe how the high-velocity blueshifted absorption features of both lines are fully consistent for each epoch. 

\paragraph{MV Lyr:} The excesses diagnostic diagram in the bottom-left panel of Fig. \ref{wings} indicates significant wind-type features in all epochs but epoch \#1. A particularly high-velocity blueshifted absorption feature in epoch \#3 (see Fig. \ref{mv_lyr}) implies a wind velocity of $\sim$2200 km s$^{-1}$ (epoch \#3a). In the bottom-left panel of Fig. \ref{overlap} we plot \ha and \hei \lam5876 for epochs \#2b and \#3a, evidencing the match between the absorption components.

\paragraph{V425 Cas:} P-Cygni-like features evident in Figs. \ref{v425_cas} and \ref{wings} (bottom-right panel) suggest the detection of a wind in epoch \#3 ($\sim$1400 km s$^{-1}$ in epoch \#3a). In the bottom-right panel of Fig. \ref{overlap}, we show an example of both the low- and \mbox{high-velocity} blueshifted absorption components of V425 Cas, visible in \hei \lam5876.

%--------------------------------------------------------------------
\section{Discussion}
We analysed optical spectra of four AWDs, namely BZ Cam, \mbox{V751 Cyg}, \mbox{MV Lyr}, and V425 Cas,  taken over three to four different observing epochs (see Table \ref{log}). All the observations were obtained during bright states, when the optical emission is expected to be dominated by the accretion flow (see Fig. \ref{lc}). We focused our analysis on the evolution of four relatively strong emission lines: H$\alpha$, \hei \lam5876, \lam6678,  and \lam7065. These transitions are known to be sensitive to the presence of wind-type outflows in different classes of stellar objects \citep[e.g.][]{Prinja1994,Kafka2004,Munoz-Darias2019}. In all four cases, the profiles of the emission lines are complex and variable  over short (i.e. same epoch: minutes) and long \mbox{(different epochs: $>$ days)} timescales. These include the presence of variable absorption components, which in some cases are observed to reach large negative velocities (i.e. blueshifted by more than \mbox{$\sim$1000 km s$^{-1}$}). These high-velocity components are not present in the red wing and are seen in all four sources studied. We tentatively identify them as signatures of wind-type outflows.

Signatures of optical outflows have previously been detected in only a handful of AWDs, including \mbox{BZ Cam} and \mbox{V751 Cyg} \citep[e.g.][]{Ringwald1998,Patterson2001,Kafka2004,Kafka2009}. Our analysis confirms these detections and adds MV Lyr and V425 Cas to the short list of systems showing features consistent with the presence of optical winds. The outflows were detected through blueshifted absorptions observed in most cases simultaneously with red emission excesses (i.e. P-Cygni profiles; see Fig. \ref{wings}). Additional features commonly present in objects with strong outflows, such as broad emission line wings \citep[e.g.][]{Munoz-Darias2016}, are also detected in both H and He emission lines. AWDs are known to exhibit strong ultraviolet winds \citep{Drew1990,Froning2005}, as is the case for BZ Cam \citep[e.g.][]{Prinja2000a}. Likewise, evidence of ultraviolet winds has been reported for V751 Cyg \citep[][]{Zellem2009} and MV Lyr \citep[][]{Linnell2005}. However, we find no records of detections for \mbox{V425 Cas}. 

In addition to high-velocity blueshifted absorption components, we observe variable and broad lower velocity absorption features (\mbox{$\lesssim~$1000 km s$^{-1}$}) underlying the main emission lines. These components can be detected on both wings of the emission lines (Figs. \ref{bz_cam} to \ref{v425_cas}) and although they are difficult to isolate from the aforementioned blueshifted absorption features that we associate with winds (and vice versa), there are clear cases where both absorption features are easily discriminated (e.g. see the \hei \lam5876 line in epoch \#3c of the upper-right panel of Fig. \ref{overlap}). It is worth noting that several AWDs with orbital periods of \mbox{3-4 h} have been classified as \mbox{SW Sex} stars \citep[e.g.][]{Rodriguez-Gil2007a,Dhillon2013}. Among other properties, this class is characterised by spectra with complex and variable line profiles, including absorption features that are believed to have an orbital dependence, swinging between positive and negative velocities \citep[e.g.][]{Casares1996}. These features, the origin of which is unknown, might be associated with the low-velocity absorption components that we see in our spectra. On the other hand, the high-velocity absorption components are only found in the blue wings, often accompanied by a red component in emission (see Fig. \ref{wings}), as one would expect if they were associated with \mbox{wind-type} outflows.

%---------------------------------------
\subsection{Observational properties of the disc winds}
We find that \ha and \hei \lam5876 are the most sensitive lines to the presence of winds. Outflow signatures at similar velocities are often observed simultaneously in both lines. This is consistent with previous observations of winds in both AWDs and XRBs \citep[e.g.][]{Kafka2003,Munoz-Darias2019}. The \hei \lam7065 line is usually a weaker version of \hei \lam5876 \citep[e.g.][]{Kafka2004}. However, in this study we only detected wind features in the \hei \lam7065 line of \mbox{BZ Cam}, the system that shows the strongest wind signatures. This system also shows strong wind signatures in \hei \lam6678, which is a singlet transition and is expected to be less sensitive to the wind under low-density conditions \citep[e.g.][]{Kafka2004}.
 
Figures \ref{bz_cam} to \ref{v425_cas} show the high variability of the wind signatures across short (minutes) and long (> day) timescales. Although our results are (mostly) in good agreement with previous studies (see below), the aforementioned variability might explain the different wind velocities reported for some cases. The highest terminal velocity (blue edge) that we measured in \mbox{BZ Cam} is \mbox{$\sim$2400 km s$^{-1}$} (see Section \ref{velocities}), which is consistent with that obtained by \citet[][\mbox{$\sim$2350 km s$^{-1}$}]{Honeycutt2013}. \citet[][]{Ringwald1998} derived velocities of up to \mbox{$\sim$3000 km s$^{-1}$} for the same system (see Section \ref{sources}). For \mbox{V751 Cyg}, we measure wind velocities of up to \mbox{$\sim$2500 km s$^{-1}$}, in agreement with the same value reported by \citet[][]{Patterson2001}. In the cases of \mbox{MV Lyr} and \mbox{V425 Cas,} our highest wind velocity measurements are \mbox{$\sim$2200 km s$^{-1}$} and \mbox{$\sim$1400 km s$^{-1}$}, respectively. All these values are consistent with those obtained by \citet[][\mbox{300--4600 km s$^{-1}$}; see their Table 4]{Kafka2004}. Likewise, ultraviolet winds with a wide range of velocities have been reported in different studies [e.g. from $\sim$165 km s$^{-1}$ for MV Lyr \citep{Linnell2005} to \mbox{$\sim$5000 km s$^{-1}$} for BZ Cam \citep{Prinja2000a}], evidencing, once again, a high variability in the observational properties of the outflows from AWDs. 

Finally, it has been suggested that the visibility of winds, both at ultraviolet and optical wavelengths, depends on the orbital phase \citep[e.g.][]{Kafka2009}. We note that this scenario was observed for BZ Cam \citep[][]{Griffith1995,Honeycutt2013} and \mbox{V751 Cyg} \citep[][]{Patterson2001}. However, \citet[][]{Patterson1996} did not observe any clear correlation between the orbital phase and the presence of optical winds in BZ Cam. Likewise, no correlation was detected in ASAS J071404$+$7004.3 \citep{Inight2022}, while the phase coverage presented by \citet[][]{Ringwald1998} and \citet[][]{Kafka2003} in BZ Cam and Q Cyg, respectively, was not wide enough (poor overlapping coverage) to test the dependency scenario. The sampling of our database and the lack of absolute ephemeris prevent us from properly testing these claims. We simply note that a correlation between orbital phase and wind signatures is not obvious in our database (e.g. systems seem to show winds signatures at different orbital phases).

%---------------------------------------
\subsection{Outflows in AWDs versus XRBs}
\label{awds_vs_xrbs}
Optical wind signatures similar to those presented here (e.g. line profiles, variability properties, and terminal velocities) have been observed in XRBs with both black hole and neutron star accretors \citep[e.g.][]{Munoz-Darias2016, Munoz-Darias2020, MataSanchez2022}. XRB winds are also observed in \mbox{X-rays} (e.g. \citealt{Ponti2012}) and in the ultraviolet (\citealt{CastroSegura2022}) and infrared (\citealt{Sanchez-Sierras2020}) regimes with similar observational properties (e.g. velocities). Indeed, they have been proposed to be multi-phase in nature (see \citealt{Munoz-Darias2022} for a discussion on this topic). The mechanism(s) behind the launch of disc winds in XRBs is still strongly debated. In this context, the similarities between the observational properties of AWDs and XRBs might be relevant. For instance, AWD ultraviolet outflows are thought to be powered by the line-driven mechanism, where ultraviolet photons are absorbed by disc ions (metals), transferring enough momentum to launch a wind \citep[e.g.][]{Drew1997,Froning2005,Honeycutt2013}. Based on the similarities between ultraviolet and optical wind features in AWDs, it is tempting to consider that this mechanism also plays a role in XRBs, a conclusion that has some theoretical support (e.g. \citealt{Higginbottom2020}). As the strong X-ray emission produced in the inner disc would over-ionise the XRB wind \citep[][]{Proga2002}, at least part of the outflow needs to be somehow shielded \citep[see e.g.][]{Motta2017,Munoz-Darias2022}. In addition, AWDs show significantly lower Eddington-scaled luminosities than those of XRBs, which poses a challenge to the scenario where both AWD and XRB disc winds have a common physical origin. 

Another interesting aspect to consider is the geometry of the wind. Given that P-Cygni profiles have been preferentially observed in low-inclination systems \citep[see, however,][]{Inight2022}, it is traditionally assumed that AWD winds are bipolar \citep[e.g.][]{Honeycutt1986,Drew1987}. In agreement with this, the systems in our sample are thought to have relatively low orbital inclinations (see Table \ref{sample}). However, most of the optical wind detections in XRBs come from high-inclination systems, although there is a growing number of systems that show winds at lower inclinations (see \citealt{Panizo-Espinar2021, Panizo-Espinar2022}). Following on from the above, it seems clear that more observations of winds in both AWDs and XRBs are needed to properly understand their geometries as well as to constrain the mechanism(s) behind these outflows.

Also, hydrodynamical modelling of line-driven winds in CVs has provided some additional insights into the geometry of disc winds. \citet{Proga1998} found that the outflow geometry might be regulated by the geometry of the radiation field. If the disc luminosity were to dominate the radiation field, a polar wind geometry would be favoured. However, if the radiation comes mainly from the compact star, the wind would be more equatorial. In addition, it is worth noting that a number of numerical simulations were able to (roughly) reproduce the spectra of high-inclination AWDs by considering biconical outflow geometries (\citealt{Knigge1997,Noebauer2010,Matthews2015,Inight2022}).

%--------------------------------------------------------------------
\section{Conclusions}
We present multi-epoch optical spectroscopy of four AWDs \mbox{(BZ Cam,} V751 Cyg, Mv Lyr, and V425 Cas). All of them show significant variability in their emission line profiles on both short (minutes) and long (days) timescales, with complex absorption and emission components. In particular, we report the detection of high-velocity ($>$ 1000 km s$^{-1}$) blueshifted absorption features that we argue can be associated with wind-type outflows. These possible wind features are detected in hydrogen (H$\alpha$) and helium emission lines (\hei \lam5876, \lam6678, and \lam7065). Our results reinforce the idea that, in addition to ultraviolet outflows, optical winds might also be common in AWDs. However, further observations are needed to fully constrain their observational properties and understand their origin and geometry, as well as their possible relation with the wind-type outflows detected in XRBs.

%______________________________________________________________
\begin{acknowledgements}
      We thank the anonymous referee for their useful and thoughtful comments, which helped to improve this paper. We acknowledge support from the Spanish Ministry of Science and Innovation via the \textit{Europa Excelencia} program EUR2021-122010 and the \textit{Proyectos de Generación de Conocimiento} PID2020-120323GB-I00 and PID2021-124879NB-I00. We acknowledge support from the \textit{Consejería de Economía, Conocimiento y Empleo del Gobierno de Canarias} and the European Regional Development Fund under grant with reference ProID2021010132 ACCISI/FEDER, UE. \\
      
      This research is based on observations made with the Gran Telescopio Canarias (GTC), installed at the Spanish Observatorio del Roque de los Muchachos of the Instituto de Astrof\'isica de Canarias, on the island of La Palma. We acknowledge with thanks the variable star observations from the AAVSO International Database contributed by observers worldwide and used in this research. \textsc{molly} software developed by Tom Marsh is gratefully acknowledged. 

      We made use of \texttt{numpy} \citep{Harris2020}, \texttt{astropy} \citep[][]{Robitaille2013,Price-Whelan2018} and \texttt{matplotlib} \citep{Hunter2007} \textsc{python} packages. We also used extensively \textsc{molly} software (\url{http://deneb.astro.warwick.ac.uk/phsaap/software/molly/html/INDEX.html}) and \textsc{iraf} \citep{Tody1986}.

\end{acknowledgements}

%-------------------------------------------------------------------
\bibliographystyle{aa} % style aa.bst
\bibliography{references}

%-------------------------------------------------------------------
\begin{appendix}
    \section{Excesses diagnostic diagram of the H$\alpha$ emission line for V425 Cas}
\begin{figure}
     \centering
     \includegraphics[trim=0mm 5mm 0mm 0mm,width=\columnwidth]{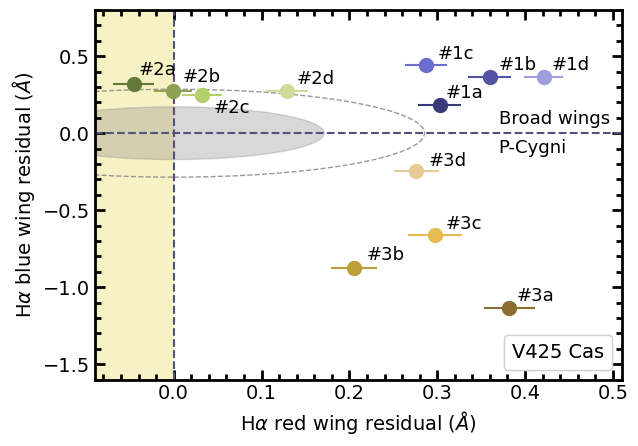}
     \caption{Excesses diagnostic diagram of \ha wings of V425 Cas using the spectral region \mbox{$\pm$400-1500 km s$^{-1}$} to estimate the residuals. The filled grey and dashed empty ellipses denote the 3$\sigma$ and 5$\sigma$ significance regions, respectively.}
     \label{Ha_diag_V425Cas}
    \end{figure}

    Figure \ref{Ha_diag_V425Cas} shows the excesses diagnostic diagram for \ha of \mbox{V425 Cas}, where residuals were calculated over the particular spectral region \mbox{$\pm$400-1500 km s$^{-1}$.} See Section \ref{wings_diagram} for more details.

\end{appendix}

\end{document}